\documentclass[longauth]{aa}
\usepackage{capt-of}
\usepackage{rotating}
\usepackage{natbib}
\usepackage{lscape}
\usepackage{color}
\usepackage{graphicx}
\newcommand{\lsun}{\mbox{$L_\odot$}}

\newcommand{\mearth}{\mbox{$M_\oplus$}}
\newcommand{\mjup}{\mbox{$M_{Jup}$}}
\newcommand{\msun}{\mbox{$M_\odot$}}

\newcommand{\mic}{\mbox{$\mu$m}}
\newcommand{\app}{\mbox{$\sim$}}
\newcommand{\sig}{\mbox{$\sigma$}}
\newcommand{\pp}{\mbox{$\pm$}}

\newcommand{\dg}{\mbox{$^\circ$}}
\newcommand{\prim}{\mbox{HD~114082}}

\newcommand{\eg}{e.g.}

\begin{document}

\title{The SHARDDS survey: First resolved image of the HD114082 debris disk in the Lower Centaurus Crux with SPHERE}
\titlerunning{First resolved rmage of the \prim\  debris disk} 
\institute{
  European Southern Observatory, Alonso de C\`{o}rdova 3107,  Vitacura, Casilla 19001, Santiago, Chile\label{inst1}\and
  Institute of Astronomy, University of Cambridge, Madingley Road, Cambridge CB3 0HA, UK\label{inst2}\and
  Steward Observatory, Department of Astronomy, University of Arizona,
  933 N. Cherry Avenue, Tucson, AZ 85721\label{inst3}\and
  LESIA, Observatoire de Paris, PSL Research Univ., CNRS, Sorbonne Univ., UPMC Univ. Paris 06, Univ. Paris Diderot, Sorbonne Paris Cité, 92195 Meudon, France\label{inst4}\and
  Instituto Nacional de Astrof\'{\i}sica, \'Optica y Electr\'onica, Luis Enrique Erro 1, Sta. Ma. Tonantzintla, Puebla, Mexico\label{inst5}\and
  UMI-FCA, CNRS/INSU, France (UMI 3386)\label{inst6}\and
  Dept. de Astronomía, Universidad de Chile, Santiago, Chile\label{inst7}\and
  Univ. Grenoble Alpes, CNRS, IPAG, F-38000 Grenoble, France\label{inst8}\and
  STAR Institute, Universit\'e de Li\`ege, 19c All\'ee du Six Ao\^ut, B-4000 Li\`ege, Belgium\label{inst9}\and
  Space Telescope Science Institute, 3700 San Martin Drive, Baltimore, MD 21218, USA\label{inst10}\and
  Hubble Fellow at Jet Propulsion Laboratory, Caltech, 4800 Oak Grove Drive, Pasadena, CA 91109, USA\label{inst11}\and
  Department of Physics, Nagoya University, Nagoya, Aichi 464-8602, Japan\label{inst12}\and
  Dept. of Astronomy, California Institute of Technology, 1200 E.\ CA Boulevard, Pasadena, CA 91125, USA\label{inst13}\and
  ALMA Santiago Central Offices, Alonso de C\`{o}rdova 3107, Vitacura, Casilla 763 0355, Santiago, Chile\label{inst14}
}

\author{
Zahed Wahhaj\inst{\ref{inst1}}\and 
Julien Milli\inst{\ref{inst1}}\and  
Grant Kennedy\inst{\ref{inst2}}\and  
Steve Ertel\inst{\ref{inst3}}\and  
Luca Matr\`a\inst{\ref{inst2}}\and
Anthony Boccaletti\inst{\ref{inst4}}\and  
Carlos del Burgo\inst{\ref{inst5}}\and   
Mark Wyatt\inst{\ref{inst2}}\and  
Christophe Pinte\inst{\ref{inst6},\ref{inst7}}\and  
Anne Marie Lagrange\inst{\ref{inst8}}\and
Olivier Absil\inst{\ref{inst9}}\and
Elodie Choquet\inst{\ref{inst11}}\and 
Carlos A.~G\'omez Gonz\'alez\inst{\ref{inst9}}\and
Hiroshi Kobayashi\inst{\ref{inst12}}\and
Dimitri Mawet\inst{\ref{inst13}}\and
David Mouillet\inst{\ref{inst8}}\and 
Laurent Pueyo\inst{\ref{inst10}}\and 
William R.F. Dent\inst{\ref{inst14}}\and
Jean-Charles Augereau\inst{\ref{inst8}}\and
Julien Girard\inst{\ref{inst1}}
}

\abstract{ We present the first resolved  image of the debris disk around the
  16\pp8~Myr old star, \prim. The  observation was made in the
  $H$ band using the SPHERE instrument. The star is at a
  distance of 92$\pm$6~pc in the Lower Centaurus Crux association.
  Using a Markov Chain Monte Carlo analysis, we determined  that
  the debris is likely in the form of a dust ring with an inner edge of
  27.7$^{+2.8}_{-3.5}$au, position angle -74.3\dg$^{+0.5}_{-1.5}$, and an inclination with respect to the
  line of sight of 6.7\dg $^{+3.8}_{-0.4}$ . The disk imaged in scattered light 
  has a surface density that is declining with radius  of \app\ r$^{-4}$, which is steeper than expected for grain
  blowout by radiation pressure. We find only marginal evidence
  (2$\sigma$) of eccentricity and rule out planets more massive than 1.0~\mjup\ 
  orbiting within 1~au of the inner edge of the ring, since such a planet would have disrupted the disk.   
  The disk has roughly the same fractional disk luminosity ($L_{disk}/L_*$=3.3$\times$10$^{-3}$) as HR~4796~A and
  $\beta$~Pictoris, however it was not detected by previous
  instrument facilities most likely because of its small angular size
  (radius$\sim$0.4$''$), low albedo (\app 0.2), and low scattering efficiency far from the star 
  due to high scattering anisotropy. 
  With the arrival of extreme adaptive optics systems, such as SPHERE and GPI, the
  morphology of smaller, fainter, and more distant debris disks are
  being revealed, providing clues to planet-disk interactions in young
  protoplanetary systems.  \\\\
 {\bf Accepted by A\&A on 7 November 2016.}
}


\maketitle

\section {Introduction}

Debris disks are dust belts produced by collisions between
planetesimals orbiting stars at ages $\gtrsim$~10~Myr
\citep[\eg,][]{1993prpl.conf.1253B,2008ARA&A..46..339W}. 
Since the first image of a debris disk around $\beta$ 
Pictoris \citep{1984Sci...226.1421S}, more than 80 debris 
disks have been resolved at optical, infrared, and submillimeter wavelengths\footnotemark \citep{2016ApJ...817L...2C}. 
Asymmetries in these dust disks are thought to be signs of
interactions with unseen bodies, possibly of planetary mass; examples include offsets of 
the disk with respect to the star as in the case of HR~4796~A (Wahhaj
et al.\ 2014\nocite{2014A&A...567A..34W}, Thalmann et al.\ 2011\nocite{2011ApJ...743L...6T}, Schneider et al.\ 2009\nocite{2009AJ....137...53S}), a warp in the
disk as in the case of the $\beta$ Pictoris planetary system \citep{2012A&A...542A..40L}, or multiple gaps and rings as in the case of HD~141569 (Perrot et al.\
2016\nocite{2016A&A...590L...7P}, Biller et al.\ 2015\nocite{2015MNRAS.450.4446B}). Indeed several debris disk systems have massive planets that
have been directly imaged, but the planet-disk interactions are not 
always well understood (\eg,\ HR 8799, Booth et al.\ 2016\nocite{2016MNRAS.460L..10B}; Fomalhaut, Kalas et al.\
2008, 2013\nocite{2008Sci...322.1345K}\nocite{2013ApJ...775...56K}; 
$\beta$ Pic, Lagrange et al.\ 2009, 2010, 2012\nocite{2009A&A...493L..21L}\nocite{2010Sci...329...57L}\nocite{2012A&A...542A..40L}; 
HD 106906, Kalas et al.\ 2015\nocite{2015ApJ...814...32K}, Lagrange et al.\ 2016\nocite{2016A&A...586L...8L}). Even toy model predictions are not easy to
make (\eg,\ Mustill \& Wyatt 2009\nocite{2009MNRAS.399.1403M}, Rodigas et al.\ 2014\nocite{2014ApJ...780...65R}). 
Moreover, significant asymmetries in the form of clumps in the
debris disk of AU~Mic have been recently attributed to a stellar wind
around the M~star primary, showing that such features need not be
connected to orbiting companions \citep{2015Natur.526..230B}. 

In order to be able to discern between different effects, it is 
essential to understand the optical properties of different grains and their
dynamical properties as an ensemble across systems of different ages and
around stars of different spectral types. 
With this goal in mind, we are undertaking the SHARDDS (SPHERE High
Angular Resolution Debris Disk Survey, VLT program 096.C-0388, PI:
J. Milli) project. Using adaptive optics (AO) imaging in $H$-band, we targeted 55 cold debris disks with high fractional
luminosity ($L_{dust}/L_* > 10^{-4}$) but that were never resolved in scattered light.
Indeed, it was unclear why highly sensitive instruments such as the Hubble Space Telescope (HST) had been
unable to detect most of these debris disks. One possibility was that the excess
emission originated from disks with small angular separations from their primaries where HST
and first generation AO instruments provided
insufficient contrasts. With the advent of an extreme adaptive optics 
instrument like the Spectro-Polarimetric High-contrast Exoplanet
REsearch (SPHERE; Beuzit et al.\ 2008\nocite{2008SPIE.7014E..18B}),
the detection of such disks is within reach.

In this paper, we present the first resolved image of a compact disk around the
F3V star \prim\ at a distance of 92\pp 6~pc with an age of 16\pp8~Myr in the Lower Centaurus
Crux association \citep{2012ApJ...746..154P}.  The luminosity and mass
estimates for the star are 3.6\pp0.2~\lsun\ and 1.4~\msun , respectively \citep{2012ApJ...746..154P}. 
Debris disks in this age range represent an interesting evolutionary
stage, as they fall between the 10~Myr old TW Hya association disks
(HR~4796A, etc;\nocite{1995ApJ...454..910S}\ Stauffer et al.\ 1995) and those in the 20~Myr old $\beta$ Pic association \citep{2014MNRAS.438L..11B}.

\footnotetext[1]{The Catalog of Circumstellar Disks: http://www.circumstellardisks.org/}

\section{Observation}
We observed \prim\ (F3V, V=8.2) on UT Feb 14, 2016 in the IRDIS \citep{2008SPIE.7018E..59D} classical imaging mode of SPHERE \citep{2008SPIE.7014E..18B} with
pupil tracking for angular difference imaging \citep{2004Sci...305.1442L,2006ApJ...641..556M}.
The SPHERE instrument is equipped with the extreme adaptive optics system SAXO \citep{2014SPIE.9148E..1UF}, which corrects 
atmospheric turbulence at 1.38~kHz, using a 40x40 lenslet
Shack-Hartmann sensor and a high order deformable mirror. The
IRDIS instrument has a field of view (FoV) of 11$'' \times$11$''$ and a pixel
scale of 12.251\pp 0.005~mas (SPHERE User manual).

We acquired 152 images, each with an integration time of 16 seconds.
The sky rotated through 16.7\dg\ with respect to the detector during this sequence.
The apodized Lyot mask, N\_ALC\_YJH\_S, with an opaque mask of diameter 185
mas was used. Images in this mode are obtained in two channels simultaneously,  both in the
H band ($\lambda$=1.625~\mic\ , $\Delta \lambda$=0.29~\mic ). Since the simultaneous 
images are in the same band, they are just added together.
The seeing ranged from 0.6$''$ to 0.7$''$ and the Strehl ratio estimated by
SAXO was 65\% to 75\%, while the wind speed varied between 1.5 and 2~m/s.

The images were flatfielded and sky subtracted in the usual way.
We also acquired off-mask unsaturated images of the star,
through a neutral density filter. These are used to estimate the
contrast achieved with respect to the star in the rest of the field.
We also acquired images with four satellite spots equidistant (400
mas) from the star, which we later used to the determine the stellar
position behind the mask. This is important when derotating images
around an accurate center. The spots are created by introducing
sine aberrations into the deformable mirror of the SAXO system.

\section{Data reduction}
\label{sec:reduc}

\begin{figure*}[ht]
  \centering
  \includegraphics[trim={0 0 0 5cm},clip,width=16cm]{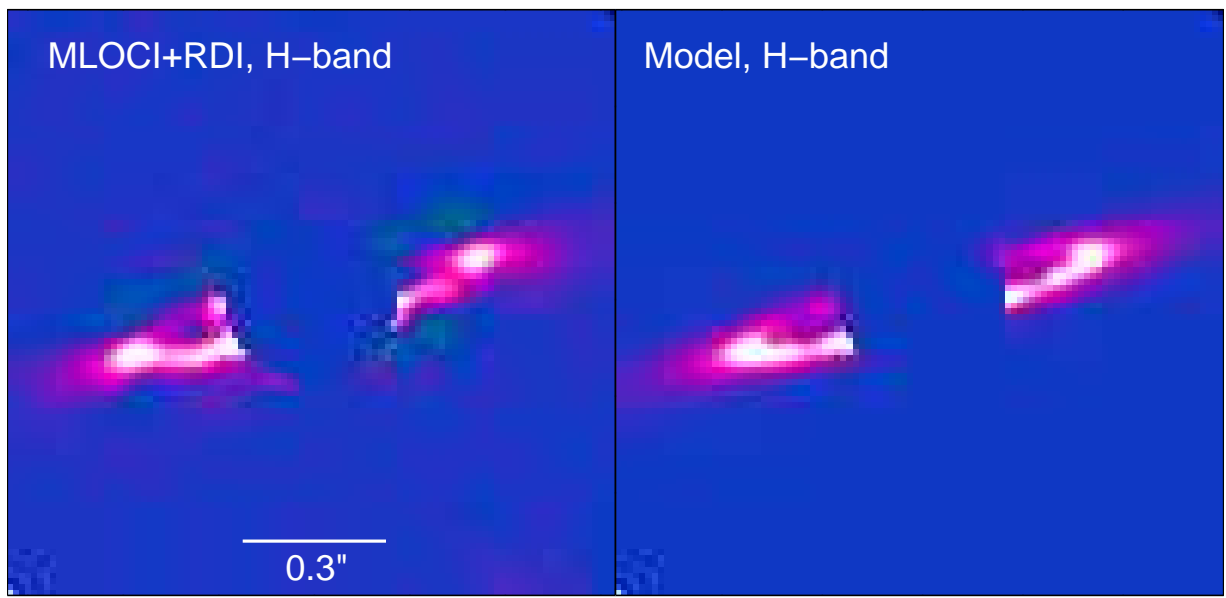}
  \caption{Left: MLOCI+ADI reduction using the
    template in Figure~\ref{mloci_rdi_temp_red}. Right: The best-fit model to MLOCI+ADI reduction
    obtained using the MCMC analysis in section 4.}
\label{mloci_red}
\end{figure*} 

Basic reduction was carried out using the SPHERE data reduction pipeline
\citep{2008ASPC..394..581P} to subtract backgrounds and apply bad pixel and flat-field corrections.

Since the total sky rotation was only 16.7\dg\ we expect significant
self-subtraction of any disk. Nonetheless, a principal component
analysis (PCA) reduction \citep{2011ApJ...741...55S} revealed an
edge-on disk extending out to \app~0.5$''$
from the star. As in any ADI reduction, we try to subtract the star
light from the science frames, which is decoupled from any
off-axis astrophysical light source that rotates. In PCA, we construct an eigenbasis 
of images using the science frames, and construct a reference
coronagraphic image using only a
few of the largest components of this image projected onto the
eigenbases. This method preferentially selects star light over the
varying (rotating) signal, thus subtraction of the reference image reveals the
underlying off-axis astrophysical source. The difference images are then
derotated and combined. 

For our reduction we found that applying the PCA algorithm to
images divided into annular rings of ten pixels in width, starting at five
pixels from the center, gave the best results. We used only the first
seven principal components to construct the reference image for subtraction.
The final reduced image is shown in Figure~\ref{pca_rdiloci}.
We see a nearly edge-on, narrow ring with a large inner
hole that is clearly visible on both sides of the star. We also performed a 
reference difference imaging (RDI) reduction, which is described in appendix~\ref{sec:rdi}.

To maximize the S/N of the disk, we use the MLOCI algorithm
\citep{2015A&A...581A..24W}. 
Here, the star subtraction and science combination steps are performed
simultaneously, while preserving  a given signal region and minimizing
the RMS in a given background region (Figure~\ref{mloci_rdi_temp_red},
right panel). 
In the reduced image of Figure~\ref{mloci_red}, we
see that the inner hole is again clearly visible and so there is a brightness asymmetry across the short axis of the ring. 
However, there is some self-subtraction although less severe than before. 

\section{Analysis}
\label{section:results}
All of our reduced images suffer from artifacts of the reduction
process. Although it is possible to estimate the range of disk
properties by  subtracting simulated disks from the data set
and repeating the reduction process to statistically analyse the
residuals, the process is very computationally expensive. 
Moreover, since two of our methods are particularly resistent to
disk self-subtraction effects, methods using simulated disks are
not critical. Given the complexity of the ring structure seen, we need to
consider nine disk parameters, a number that is very large for 
a Markov Chain Monte Carlo (MCMC) analysis.
These model parameters are the following: (1) dust surface density, $\Sigma_0$; 
(2, 3) offsets of the disk from the star, $\vec{x}$ and $\vec{y}$;
(4) inclination to the line of sight (LOS), $\phi$
(0\dg\ indicating a pole-on orientation); (5) position angle (PA),
measured east of north; (6) inner
radius of ring, $r_{in}$; (7) ring width, $\Delta
r$ ; (8) ring tail, exponent of density profile, $\gamma$;  and (9) Henyey-Greenstein scattering parameter, $g$.

We model the dust surface density as $\Sigma(r) =  0$ for $r < r_{in}$, 
 $\Sigma(r) =  \Sigma_0$ for  $r\ =\ r_{in}\ to\ r_{in} + \Delta r $,
 $\Sigma(r) =  \Sigma_0\ r^{-\gamma}$ for $r > r_{in} + \Delta r $. 
Here, $r$ is the distance from the star. The
contrast of the disk is modeled as $f(\vec{R})/f_* = p(\theta) \omega \Sigma(r)/{\|\vec{R}\|}^2$,
where $\vec{R}=\vec{r}+\vec{x}+\vec{y}$, $f_*$ is the flux from the star and $\omega$ is the albedo.
We can only constrain the product $\omega \Sigma_0$ now, but we break this degeneracy by considering
infrared flux measurements later. 
The flat density segment (from $\ r_{in}\ to\ r_{in} + \Delta r $) is included to emulate a planetesimal ``birth'' ring, where
large grains mostly unaffected by radiation pressure maintain their orbits.
Beyond this radius,  the density is described by a power law to
model the distribution of small grains being blown out. 
Anisotropic scattering is described by the Henyey-Greenstein function,
$p(\theta)={(1-g^2)}/{(4\pi(1 + g^2 - 2g cos(\theta)))^{3/2}}$, where
$\theta$ is the angle wrt.\ the LOS. 

As in \citet{2014A&A...567A..34W}, we use the Metropolis Hastings
measure with MCMC to generate a sampling of the nine-dimensional
parameter space, which is also an estimate of the relative probability distribution for
the space; see that article for details. The relative probability of a model is
$e^{-{\chi^2}/2}$, where $\chi^2$ is the usual statistic for
data and model image comparisons.  The
model and data are compared over an ellipse (aligned with the disk) of major axis
2$''$ and minor axis 0.13$''$, with the central circular region of
radius 0.13$''$ excluded. This region includes all regions of the disk
that are significantly detected and an ample background region,
but excludes regions containing mostly stellar residuals. 
Including background regions is important for constraining the edges of the ring.
In total, it covers 226 resolution elements. However, we rescale this to count only 76
independent measurements that register above a detection level of 2$\sigma$. 
The flux uncertainty per pixel, used to calculate
$\chi^2$, is the standard deviation of intensities in the matching
region, but with the disk effectively removed by a filtering
process. The technique is similar to that used in \citet{2013ApJ...779...80W},
except that it removes running flux averages over 15 pixels
along the radial (instead of the azimuthal) direction with the star as center. 
To ensure good constraints, we examined the probability distribution of models from
MCMC to make sure that the probability of each parameter agrees to within
10\% over two different runs.   
Along with the best-fit parameter values, we list the 95\% confidence internal 
for each parameter by rejecting 2.5\% of the values on both extremes of the distribution (see Table~\ref{tab:ringprams}).   
Using the contrast measurement of the disk ansae from the LOCI+RDI
reduction (appendix~\ref{sec:rdi}), we  estimate $\omega \Sigma_0$
$=$3.28\pp0.14$\times$10$^{-3}$, which sets the overall intensity
scale.

The parameter estimates obtained from the fits to the different reduced images are 
consistent in most cases, but some have non-overlapping confidence intervals with small fractional differences in the best estimates. 
This is true for the ring inclination where we have a \app\ 5\%
difference (81\dg\ vs 84.9\dg ) and the inner radius of the ring with \app\ 12\% difference (25~au versus 28.5~au).
Also, the PCA reduction is insensitive to scattering anisotropy due
to disk self-subtraction and so we ignore its constraints on $g$.
We believe that the non-overlapping constraints are due to small artifacts introduced into the final images from the reduction
process.  Since the MLOCI reduced image yields the smallest reduced $\chi^2$ (1.5), we adopt its parameter estimates as best, while
reporting the extremes of the 2$\sigma$ limits from the other reductions to be conservative in our constraints.  

We repeat the MCMC analysis, comparing the spectral energy distribution (SED) models to the available
photometry (Table~\ref{tab:phot}) with $a_0$ (minimum grain radius), $q$
(grain size distribution exponent), $\omega \Sigma_0$ as free
parameters (see appendix~\ref{sec:sed}).
Meanwhile, $r_{in}$,  $\gamma$, and $\Delta r$ are allowed to vary within the extremes of the
ranges permitted by the image fits (see Table~\ref{tab:ringprams}). 
For all models, the total dust mass $m_d$ in grains smaller than $4$ mm
\citep[see][]{2012A&A...541A.148E} is calculated assuming a density of  2300~$kg/m^3$.
The 2$\sigma$ constraints found were $a_0 =$\ 5.0--18~\mic\ and $q =$\ 3.9--7.8, while 
$M_d =$\ 0.022-0.043~\mearth . Using the previous imaging constraints on
$\omega \Sigma_0$ with the SED constraint on $(1-\omega) \Sigma_0$, we
find the albedo, $\omega=$0.13--0.24. See Figure~\ref{sed} for the
best SED model fit to the photometry.
Finally, the AKARI photometry suggests that there is excess emission at 9~\mic\ 
coming from warmer dust than is detected in our scattered light images.  We can explain this excess with
a ring extending from 3--4~au with the same surface density as the birth ring. Of course this solution is 
not unique.  


\section{Discussion and conclusions}
\label{sec:discuss}
We have presented the first resolved image of the debris disk around \prim , 
a 16\pp8~Myr old, F3V star in the Lower Centaurus Crux association. 
We estimate from the MCMC analysis of\ this $H$-band image that the disk has a birth ring with uniform density of width 1.9$^{+5.8}_{-0.9}$~au, outside of which the density 
falls off with a power-law index of 3.9$^{+3.3}_{-1.1}$. This is the first
detection of a debris disk from the SHARDDS program, which aimed
to explain why some disks with estimated high fractional  
luminosities were not detected in scattered light.

\prim\ joins a family of debris ring systems that have been imaged 
with clear inner holes, for example,\ HD 181327 \citep{2006ApJ...650..414S}, HD 207129 \citep{2010AJ....140.1051K}, 
HD 202628 \citep{2012AJ....144...45K}, HR~4796~A \citep{2014A&A...567A..34W,2015A&A...577A..57M,2015ApJ...799..182P}, 
and HD 106906 \citep{2015ApJ...814...32K,2016A&A...586L...8L}. Moreover, it is 
much like HR~4796~A, in that it also has significant scattering anisotropy ($g=0.23^{+0.05}_{-0.08}$), 
and has mid-sized grains (radius \app~11~\mic ), indicating that their grain
properties may be very similar.  The blowout grain size due to radiation
pressure is given by  $D = 2300/\rho\  (L_*/L_{\odot})$~$(M_{\odot}/M_*)$
\mic. The blowout size estimate for \prim\ is thus 2.4~\mic , which is much
smaller than our estimated minimum grain size  (\app~10~\mic ).
The estimated dust density power-law index (\app 3.9) is much 
steeper than what is expected for a distribution dominated by radiation
pressure blowout \citep[1.5;][]{2008A&A...481..713T}. The grain size power-law exponent ($>$ 3.9) is steeper than that of other debris disks but
is still consistent with realistic collisional cascades (see discussion in \nocite{2016ApJ...823...79M}MacGregor et al. 2016). 
Since the ring is relatively narrow, inner and outer shepherding planets may be necessary to maintain it. 
Although the  disk center is not detectibly offset from the star along its apparent long axis, because of the high 
inclination of the disk, we can only constrain the offset along the short
axis to $<$0.15$''$. This would allow a brightness asymmetry of
\app~30\% , accounting for the fact that the
density of grains is smaller at the pericenter in a Keplerian disk \citep[see][]{2014A&A...567A..34W}. 
Thus, any such eccentricity or offset would be unable to explain the
estimated brightness asymmetry factor (4 for $g$=0.23) along the short axis, which in turn strengthens
the scattering asymmetry claim. On the other hand, if we assume that the
\app 2$\sigma$ level hint of an offset (\app 0.019$"$; see Table~\ref{tab:ringprams}) along the long axis is
real, the eccentricity would be \app\ 0.02. 

The ring has a fractional width between 0.07--0.175. Consulting equations 2 \& 5
from \citep{2014ApJ...780...65R}, we find that a putative planet interior
of the disk should not be more massive than 1.0~\mjup\ to have the desired
broadening effect on the ring and should have an orbital radius of \app\ 25~au. 
These constraints are more stringent that those obtained from the
direct detection limits (see appendix~\ref{sec:conlim}).

Lastly, the non-detection of \prim\ by past facilities despite having a large fractional infrared luminosity 
is likely due to three reasons:
(1) a small disk radius (25--30~au) compared to other ring systems ($>$70\ au),
(2) a low albedo (\app 0.2), and 
(3) a relatively high scattering anisotropy ($g$\app0.23) that forces
most of the light to be scattered behind the coronagraphic mask.
Nevertheless, we expect more detections of narrow ring systems with smaller disks by extreme AO instruments like SPHERE and GPI, as 
such systems are predicted from dynamical analyses investigating the infrared excesses detected around young stars \citep{2014MNRAS.442.3266K}. 
  

\begin{acknowledgements}  
We would like to thank the ESO staff and the technical operators at the Paranal Observatory.
OA is F.R.S.-FNRS Research Associate. OA and CGG acknowledge support
by the European Union through ERC grant number 337569. 
GMK is supported by the Royal Society as a Royal Society University
Research Fellow.  CB has been supported by Mexican CONACyT research grant CB-2012-183007.
EC acknowledges support for this work from NASA through Hubble
Fellowship grant HST-HF2-51355 awarded by STScI, which is operated by
the AURA, Inc., for NASA under contract NAS5-26555.
\end{acknowledgements} 
\bibliographystyle{aa}
\bibliography{zrefs}

\begin{thebibliography}{53}
\expandafter\ifx\csname natexlab\endcsname\relax\def\natexlab#1{#1}\fi

\bibitem[{{Allard} {et~al.}(2001){Allard}, {Hauschildt}, {Alexander},
  {Tamanai}, \& {Schweitzer}}]{2001ApJ...556..357A}
{Allard}, F., {Hauschildt}, P.~H., {Alexander}, D.~R., {et~al.} 2001, \apj,
  556, 357

\bibitem[{{Backman} \& {Paresce}(1993)}]{1993prpl.conf.1253B}
{Backman}, D.~E. \& {Paresce}, F. 1993, in Protostars and Planets III, ed.
  E.~H. {Levy} \& J.~I. {Lunine}, 1253--1304

\bibitem[{{Backman} {et~al.}(1992){Backman}, {Witteborn}, \&
  {Gillett}}]{1992ApJ...385..670B}
{Backman}, D.~E., {Witteborn}, F.~C., \& {Gillett}, F.~C. 1992, \apj, 385, 670

\bibitem[{{Beuzit} {et~al.}(2008){Beuzit}, {Feldt}, {Dohlen}, {Mouillet},
  {Puget}, {Wildi}, {Abe}, {Antichi}, {Baruffolo}, {Baudoz}, {Boccaletti},
  {Carbillet}, {Charton}, {Claudi}, {Downing}, {Fabron}, {Feautrier},
  {Fedrigo}, {Fusco}, {Gach}, {Gratton}, {Henning}, {Hubin}, {Joos}, {Kasper},
  {Langlois}, {Lenzen}, {Moutou}, {Pavlov}, {Petit}, {Pragt}, {Rabou}, {Rigal},
  {Roelfsema}, {Rousset}, {Saisse}, {Schmid}, {Stadler}, {Thalmann}, {Turatto},
  {Udry}, {Vakili}, \& {Waters}}]{2008SPIE.7014E..18B}
{Beuzit}, J.-L., {Feldt}, M., {Dohlen}, K., {et~al.} 2008, in \procspie, Vol.
  7014, Instrumentation for Astronomy II, 701418

\bibitem[{{Biller} {et~al.}(2015){Biller}, {Liu}, {Rice}, {Wahhaj}, {Nielsen},
  {Hayward}, {Kuchner}, {Close}, {Chun}, {Ftaclas}, \&
  {Toomey}}]{2015MNRAS.450.4446B}
{Biller}, B.~A., {Liu}, M.~C., {Rice}, K., {et~al.} 2015, \mnras, 450, 4446

\bibitem[{{Binks} \& {Jeffries}(2014)}]{2014MNRAS.438L..11B}
{Binks}, A.~S. \& {Jeffries}, R.~D. 2014, \mnras, 438, L11

\bibitem[{{Boccaletti} {et~al.}(2015){Boccaletti}, {Thalmann}, {Lagrange},
  {Janson}, {Augereau}, {Schneider}, {Milli}, {Grady}, {Debes}, {Langlois},
  {Mouillet}, {Henning}, {Dominik}, {Maire}, {Beuzit}, {Carson}, {Dohlen},
  {Engler}, {Feldt}, {Fusco}, {Ginski}, {Girard}, {Hines}, {Kasper}, {Mawet},
  {M{\'e}nard}, {Meyer}, {Moutou}, {Olofsson}, {Rodigas}, {Sauvage},
  {Schlieder}, {Schmid}, {Turatto}, {Udry}, {Vakili}, {Vigan}, {Wahhaj}, \&
  {Wisniewski}}]{2015Natur.526..230B}
{Boccaletti}, A., {Thalmann}, C., {Lagrange}, A.-M., {et~al.} 2015, \nat, 526,
  230

\bibitem[{{Booth} {et~al.}(2016){Booth}, {Jord{\'a}n}, {Casassus}, {Hales},
  {Dent}, {Faramaz}, {Matr{\`a}}, {Barkats}, {Brahm}, \&
  {Cuadra}}]{2016MNRAS.460L..10B}
{Booth}, M., {Jord{\'a}n}, A., {Casassus}, S., {et~al.} 2016, \mnras, 460, L10

\bibitem[{{Chen} {et~al.}(2014){Chen}, {Mittal}, {Kuchner}, {Forrest}, {Lisse},
  {Manoj}, {Sargent}, \& {Watson}}]{2014ApJS..211...25C}
{Chen}, C.~H., {Mittal}, T., {Kuchner}, M., {et~al.} 2014, \apjs, 211, 25

\bibitem[{{Choquet} {et~al.}(2016){Choquet}, {Perrin}, {Chen}, {Soummer},
  {Pueyo}, {Hagan}, {Gofas-Salas}, {Rajan}, {Golimowski}, {Hines}, {Schneider},
  {Mazoyer}, {Augereau}, {Debes}, {Stark}, {Wolff}, {N'Diaye}, \&
  {Hsiao}}]{2016ApJ...817L...2C}
{Choquet}, {\'E}., {Perrin}, M.~D., {Chen}, C.~H., {et~al.} 2016, \apjl, 817,
  L2

\bibitem[{{Cutri} {et~al.}(2003){Cutri}, {Skrutskie}, {van Dyk}, {Beichman},
  {Carpenter}, {Chester}, {Cambresy}, {Evans}, {Fowler}, {Gizis}, {Howard},
  {Huchra}, {Jarrett}, {Kopan}, {Kirkpatrick}, {Light}, {Marsh}, {McCallon},
  {Schneider}, {Stiening}, {Sykes}, {Weinberg}, {Wheaton}, {Wheelock}, \&
  {Zacarias}}]{2003tmc..book.....C}
{Cutri}, R.~M., {Skrutskie}, M.~F., {van Dyk}, S., {et~al.} 2003, {2MASS All
  Sky Catalog of point sources.}

\bibitem[{{Dohlen} {et~al.}(2008){Dohlen}, {Saisse}, {Origne}, {Moreaux},
  {Fabron}, {Zamkotsian}, {Lanzoni}, \& {Lemarquis}}]{2008SPIE.7018E..59D}
{Dohlen}, K., {Saisse}, M., {Origne}, A., {et~al.} 2008, in \procspie, Vol.
  7018, Telescopes and Instrumentation, 701859

\bibitem[{{Dohnanyi}(1969)}]{1969JGR....74.2531D}
{Dohnanyi}, J.~S. 1969, \jgr, 74, 2531

\bibitem[{{Ertel} {et~al.}(2012){Ertel}, {Wolf}, {Marshall}, {Eiroa},
  {Augereau}, {Krivov}, {L{\"o}hne}, {Absil}, {Ardila}, {Ar{\'e}valo}, {Bayo},
  {Bryden}, {del Burgo}, {Greaves}, {Kennedy}, {Lebreton}, {Liseau},
  {Maldonado}, {Montesinos}, {Mora}, {Pilbratt}, {Sanz-Forcada}, {Stapelfeldt},
  \& {White}}]{2012A&A...541A.148E}
{Ertel}, S., {Wolf}, S., {Marshall}, J.~P., {et~al.} 2012, \aap, 541, A148

\bibitem[{{Fusco} {et~al.}(2014){Fusco}, {Sauvage}, {Petit}, {Costille},
  {Dohlen}, {Mouillet}, {Beuzit}, {Kasper}, {Suarez}, {Soenke}, {Fedrigo},
  {Downing}, {Baudoz}, {Sevin}, {Perret}, {Barrufolo}, {Salasnich}, {Puget},
  {Feautrier}, {Rochat}, {Moulin}, {Deboulb{\'e}}, {Hugot}, {Vigan}, {Mawet},
  {Girard}, \& {Hubin}}]{2014SPIE.9148E..1UF}
{Fusco}, T., {Sauvage}, J.-F., {Petit}, C., {et~al.} 2014, in \procspie, Vol.
  9148, Adaptive Optics Systems IV, 91481U

\bibitem[{Greenberg(1979)}]{greenberg79}
Greenberg, J.~M. 1979, in Infrared Astronomy, Proceedings of NATO Advanced
  Study Institute, (ed. G Setti and G.G. Fazio, 51)

\bibitem[{{Ishihara} {et~al.}(2010){Ishihara}, {Onaka}, {Kataza}, {Salama},
  {Alfageme}, {Cassatella}, {Cox}, {Garc{\'{\i}}a-Lario}, {Stephenson},
  {Cohen}, {Fujishiro}, {Fujiwara}, {Hasegawa}, {Ita}, {Kim}, {Matsuhara},
  {Murakami}, {M{\"u}ller}, {Nakagawa}, {Ohyama}, {Oyabu}, {Pyo}, {Sakon},
  {Shibai}, {Takita}, {Tanab{\'e}}, {Uemizu}, {Ueno}, {Usui}, {Wada},
  {Watarai}, {Yamamura}, \& {Yamauchi}}]{2010AA...514A...1I}
{Ishihara}, D., {Onaka}, T., {Kataza}, H., {et~al.} 2010, \aap, 514, A1

\bibitem[{{Kalas} {et~al.}(2008){Kalas}, {Graham}, {Chiang}, {Fitzgerald},
  {Clampin}, {Kite}, {Stapelfeldt}, {Marois}, \& {Krist}}]{2008Sci...322.1345K}
{Kalas}, P., {Graham}, J.~R., {Chiang}, E., {et~al.} 2008, Science, 322, 1345

\bibitem[{{Kalas} {et~al.}(2013){Kalas}, {Graham}, {Fitzgerald}, \&
  {Clampin}}]{2013ApJ...775...56K}
{Kalas}, P., {Graham}, J.~R., {Fitzgerald}, M.~P., {et~al.} 2013, \apj, 775, 56

\bibitem[{{Kalas} {et~al.}(2015){Kalas}, {Rajan}, {Wang}, {Millar-Blanchaer},
  {Duchene}, {Chen}, {Fitzgerald}, {Dong}, {Graham}, {Patience}, {Macintosh},
  {Murray-Clay}, {Matthews}, {Rameau}, {Marois}, {Chilcote}, {De Rosa},
  {Doyon}, {Draper}, {Lawler}, {Ammons}, {Arriaga}, {Bulger}, {Cotten},
  {Follette}, {Goodsell}, {Greenbaum}, {Hibon}, {Hinkley}, {Hung}, {Ingraham},
  {Konapacky}, {Lafreniere}, {Larkin}, {Long}, {Maire}, {Marchis}, {Metchev},
  {Morzinski}, {Nielsen}, {Oppenheimer}, {Perrin}, {Pueyo}, {Rantakyr{\"o}},
  {Ruffio}, {Saddlemyer}, {Savransky}, {Schneider}, {Sivaramakrishnan},
  {Soummer}, {Song}, {Thomas}, {Vasisht}, {Ward-Duong}, {Wiktorowicz}, \&
  {Wolff}}]{2015ApJ...814...32K}
{Kalas}, P.~G., {Rajan}, A., {Wang}, J.~J., {et~al.} 2015, \apj, 814, 32

\bibitem[{{Kobayashi} \& {L{\"o}hne}(2014)}]{2014MNRAS.442.3266K}
{Kobayashi}, H. \& {L{\"o}hne}, T. 2014, \mnras, 442, 3266

\bibitem[{{Krist} {et~al.}(2012){Krist}, {Stapelfeldt}, {Bryden}, \&
  {Plavchan}}]{2012AJ....144...45K}
{Krist}, J.~E., {Stapelfeldt}, K.~R., {Bryden}, G., {et~al.} 2012, \aj, 144, 45

\bibitem[{{Krist} {et~al.}(2010){Krist}, {Stapelfeldt}, {Bryden}, {Rieke},
  {Su}, {Chen}, {Beichman}, {Hines}, {Rebull}, {Tanner}, {Trilling}, {Clampin},
  \& {G{\'a}sp{\'a}r}}]{2010AJ....140.1051K}
{Krist}, J.~E., {Stapelfeldt}, K.~R., {Bryden}, G., {et~al.} 2010, \aj, 140,
  1051

\bibitem[{{Lagrange} {et~al.}(2009){Lagrange}, {Gratadour}, {Chauvin}, {Fusco},
  {Ehrenreich}, {Mouillet}, {Rousset}, {Rouan}, {Allard}, {Gendron}, {Charton},
  {Mugnier}, {Rabou}, {Montri}, \& {Lacombe}}]{2009A&A...493L..21L}
{Lagrange}, A., {Gratadour}, D., {Chauvin}, G., {et~al.} 2009, \aap, 493, L21

\bibitem[{{Lagrange} {et~al.}(2012){Lagrange}, {Boccaletti}, {Milli},
  {Chauvin}, \& {Others}}]{2012A&A...542A..40L}
{Lagrange}, A.-M., {Boccaletti}, A., {Milli}, J., {et~al.} 2012, \aap, 542, A40

\bibitem[{{Lagrange} {et~al.}(2010){Lagrange}, {Bonnefoy}, {Chauvin}, {Apai},
  {Ehrenreich}, {Boccaletti}, {Gratadour}, {Rouan}, {Mouillet}, {Lacour}, \&
  {Kasper}}]{2010Sci...329...57L}
{Lagrange}, A.-M., {Bonnefoy}, M., {Chauvin}, G., {et~al.} 2010, Science, 329,
  57

\bibitem[{{Lagrange} {et~al.}(2016){Lagrange}, {Langlois}, {Gratton}, {Maire},
  {Milli}, {Olofsson}, {Vigan}, {Bailey}, {Mesa}, {Chauvin}, {Boccaletti},
  {Galicher}, {Girard}, {Bonnefoy}, {Samland}, {Menard}, {Henning},
  {Kenworthy}, {Thalmann}, {Beust}, {Beuzit}, {Brandner}, {Buenzli},
  {Cheetham}, {Janson}, {le Coroller}, {Lannier}, {Mouillet}, {Peretti},
  {Perrot}, {Salter}, {Sissa}, {Wahhaj}, {Abe}, {Desidera}, {Feldt}, {Madec},
  {Perret}, {Petit}, {Rabou}, {Soenke}, \& {Weber}}]{2016A&A...586L...8L}
{Lagrange}, A.-M., {Langlois}, M., {Gratton}, R., {et~al.} 2016, \aap, 586, L8

\bibitem[{{Lieman-Sifry} {et~al.}(2016){Lieman-Sifry}, {Hughes}, {Carpenter},
  {Gorti}, {Hales}, \& {Flaherty}}]{2016ApJ...828...25L}
{Lieman-Sifry}, J., {Hughes}, A.~M., {Carpenter}, J.~M., {et~al.} 2016, \apj,
  828, 25

\bibitem[{{Liu}(2004)}]{2004Sci...305.1442L}
{Liu}, M.~C. 2004, Science, 305, 1442

\bibitem[{{L{\"o}hne} {et~al.}(2012){L{\"o}hne}, {Augereau}, {Ertel},
  {Marshall}, {Eiroa}, {Mora}, {Absil}, {Stapelfeldt}, {Th{\'e}bault}, {Bayo},
  {Del Burgo}, {Danchi}, {Krivov}, {Lebreton}, {Letawe}, {Magain}, {Maldonado},
  {Montesinos}, {Pilbratt}, {White}, \& {Wolf}}]{2012A&A...537A.110L}
{L{\"o}hne}, T., {Augereau}, J.-C., {Ertel}, S., {et~al.} 2012, \aap, 537, A110

\bibitem[{{MacGregor} {et~al.}(2016){MacGregor}, {Wilner}, {Chandler}, {Ricci},
  {Maddison}, {Cranmer}, {Andrews}, {Hughes}, \&
  {Steele}}]{2016ApJ...823...79M}
{MacGregor}, M.~A., {Wilner}, D.~J., {Chandler}, C., {et~al.} 2016, \apj, 823,
  79

\bibitem[{{Marois} {et~al.}(2006){Marois}, {Lafreni{\`e}re}, {Doyon},
  {Macintosh}, \& {Nadeau}}]{2006ApJ...641..556M}
{Marois}, C., {Lafreni{\`e}re}, D., {Doyon}, R., {et~al.} 2006, \apj, 641, 556

\bibitem[{{Milli} {et~al.}(2015){Milli}, {Mawet}, {Pinte}, {Lagrange},
  {Mouillet}, {Girard}, {Augereau}, {De Boer}, {Pueyo}, \&
  {Choquet}}]{2015A&A...577A..57M}
{Milli}, J., {Mawet}, D., {Pinte}, C., {et~al.} 2015, \aap, 577, A57

\bibitem[{{Milli} {et~al.}(2012){Milli}, {Mouillet}, {Lagrange}, {Boccaletti},
  {Mawet}, {Chauvin}, \& {Bonnefoy}}]{2012A&A...545A.111M}
{Milli}, J., {Mouillet}, D., {Lagrange}, A.-M., {et~al.} 2012, \aap, 545, A111

\bibitem[{{Mustill} \& {Wyatt}(2009)}]{2009MNRAS.399.1403M}
{Mustill}, A.~J. \& {Wyatt}, M.~C. 2009, \mnras, 399, 1403

\bibitem[{{Pavlov} {et~al.}(2008){Pavlov}, {Feldt}, \&
  {Henning}}]{2008ASPC..394..581P}
{Pavlov}, A., {Feldt}, M., \& {Henning}, T. 2008, in Astronomical Data Analysis
  Software and Systems XVII, ed. R.~W. {Argyle}, P.~S. {Bunclark}, \& J.~R.
  {Lewis}, Vol. 394, 581

\bibitem[{{Pecaut} {et~al.}(2012){Pecaut}, {Mamajek}, \&
  {Bubar}}]{2012ApJ...746..154P}
{Pecaut}, M.~J., {Mamajek}, E.~E., \& {Bubar}, E.~J. 2012, \apj, 746, 154

\bibitem[{{Perrin} {et~al.}(2015){Perrin}, {Duchene}, {Millar-Blanchaer},
  {Fitzgerald}, {Graham}, {Wiktorowicz}, {Kalas}, {Macintosh}, {Bauman},
  {Cardwell}, {Chilcote}, {De Rosa}, {Dillon}, {Doyon}, {Dunn}, {Erikson},
  {Gavel}, {Goodsell}, {Hartung}, {Hibon}, {Ingraham}, {Kerley}, {Konapacky},
  {Larkin}, {Maire}, {Marchis}, {Marois}, {Mittal}, {Morzinski}, {Oppenheimer},
  {Palmer}, {Patience}, {Poyneer}, {Pueyo}, {Rantakyr{\"o}}, {Sadakuni},
  {Saddlemyer}, {Savransky}, {Soummer}, {Sivaramakrishnan}, {Song}, {Thomas},
  {Wallace}, {Wang}, \& {Wolff}}]{2015ApJ...799..182P}
{Perrin}, M.~D., {Duchene}, G., {Millar-Blanchaer}, M., {et~al.} 2015, \apj,
  799, 182

\bibitem[{{Perrot} {et~al.}(2016){Perrot}, {Boccaletti}, {Pantin}, {Augereau},
  {Lagrange}, {Galicher}, {Maire}, {Mazoyer}, {Milli}, {Rousset}, {Gratton},
  {Bonnefoy}, {Brandner}, {Buenzli}, {Langlois}, {Lannier}, {Mesa}, {Peretti},
  {Salter}, {Sissa}, {Chauvin}, {Desidera}, {Feldt}, {Vigan}, {Di Folco},
  {Dutrey}, {P{\'e}ricaud}, {Baudoz}, {Benisty}, {De Boer}, {Garufi}, {Girard},
  {Menard}, {Olofsson}, {Quanz}, {Mouillet}, {Christiaens}, {Casassus},
  {Beuzit}, {Blanchard}, {Carle}, {Fusco}, {Giro}, {Hubin}, {Maurel},
  {Moeller-Nilsson}, {Sevin}, \& {Weber}}]{2016A&A...590L...7P}
{Perrot}, C., {Boccaletti}, A., {Pantin}, E., {et~al.} 2016, \aap, 590, L7

\bibitem[{{Rodigas} {et~al.}(2014){Rodigas}, {Malhotra}, \&
  {Hinz}}]{2014ApJ...780...65R}
{Rodigas}, T.~J., {Malhotra}, R., \& {Hinz}, P.~M. 2014, \apj, 780, 65

\bibitem[{{Schneider} {et~al.}(2006){Schneider}, {Silverstone}, {Hines},
  {Augereau}, {Pinte}, {M{\'e}nard}, {Krist}, {Clampin}, {Grady}, {Golimowski},
  {Ardila}, {Henning}, {Wolf}, \& {Rodmann}}]{2006ApJ...650..414S}
{Schneider}, G., {Silverstone}, M.~D., {Hines}, D.~C., {et~al.} 2006, \apj,
  650, 414

\bibitem[{{Schneider} {et~al.}(2009){Schneider}, {Weinberger}, {Becklin},
  {Debes}, \& {Smith}}]{2009AJ....137...53S}
{Schneider}, G., {Weinberger}, A.~J., {Becklin}, E.~E., {et~al.} 2009, \aj,
  137, 53

\bibitem[{{Smith} \& {Terrile}(1984)}]{1984Sci...226.1421S}
{Smith}, B.~A. \& {Terrile}, R.~J. 1984, Science, 226, 1421

\bibitem[{{Soummer} {et~al.}(2011){Soummer}, {Brendan Hagan}, {Pueyo},
  {Thormann}, {Rajan}, \& {Marois}}]{2011ApJ...741...55S}
{Soummer}, R., {Brendan Hagan}, J., {Pueyo}, L., {et~al.} 2011, \apj, 741, 55

\bibitem[{{Stauffer} {et~al.}(1995){Stauffer}, {Hartmann}, \& {Barrado y
  Navascues}}]{1995ApJ...454..910S}
{Stauffer}, J.~R., {Hartmann}, L.~W., \& {Barrado y Navascues}, D. 1995, \apj,
  454, 910

\bibitem[{{Thalmann} {et~al.}(2011){Thalmann}, {Janson}, {Buenzli}, {Brandt},
  {Wisniewski}, {Moro-Mart{\'{\i}}n}, {Usuda}, {Schneider}, {Carson},
  {McElwain}, {Grady}, {Goto}, {Abe}, {Brandner}, {Dominik}, {Egner}, {Feldt},
  {Fukue}, {Golota}, {Guyon}, {Hashimoto}, {Hayano}, {Hayashi}, {Hayashi},
  {Henning}, {Hodapp}, {Ishii}, {Iye}, {Kandori}, {Knapp}, {Kudo}, {Kusakabe},
  {Kuzuhara}, {Matsuo}, {Miyama}, {Morino}, {Nishimura}, {Pyo}, {Serabyn},
  {Suto}, {Suzuki}, {Takahashi}, {Takami}, {Takato}, {Terada}, {Tomono},
  {Turner}, {Watanabe}, {Yamada}, {Takami}, \& {Tamura}}]{2011ApJ...743L...6T}
{Thalmann}, C., {Janson}, M., {Buenzli}, E., {et~al.} 2011, \apjl, 743, L6

\bibitem[{{Th{\'e}bault} \& {Wu}(2008)}]{2008A&A...481..713T}
{Th{\'e}bault}, P. \& {Wu}, Y. 2008, \aap, 481, 713

\bibitem[{{Wahhaj} {et~al.}(2015){Wahhaj}, {Cieza}, {Mawet}, {Yang}, {Canovas},
  {de Boer}, {Casassus}, {M{\'e}nard}, {Schreiber}, {Liu}, {Biller}, {Nielsen},
  \& {Hayward}}]{2015A&A...581A..24W}
{Wahhaj}, Z., {Cieza}, L.~A., {Mawet}, D., {et~al.} 2015, \aap, 581, A24

\bibitem[{{Wahhaj} {et~al.}(2005){Wahhaj}, {Koerner}, {Backman}, {Werner},
  {Serabyn}, {Ressler}, \& {Lis}}]{2005ApJ...618..385W}
{Wahhaj}, Z., {Koerner}, D.~W., {Backman}, D.~E., {et~al.} 2005, \apj, 618, 385

\bibitem[{{Wahhaj} {et~al.}(2013){Wahhaj}, {Liu}, {Biller}, {Nielsen}, {Close},
  {Hayward}, \& {Others}}]{2013ApJ...779...80W}
{Wahhaj}, Z., {Liu}, M.~C., {Biller}, B.~A., {et~al.} 2013, \apj, 779, 80

\bibitem[{{Wahhaj} {et~al.}(2014){Wahhaj}, {Liu}, {Biller}, {Nielsen},
  {Hayward}, {Kuchner}, {Close}, {Chun}, {Ftaclas}, \&
  {Toomey}}]{2014A&A...567A..34W}
{Wahhaj}, Z., {Liu}, M.~C., {Biller}, B.~A., {et~al.} 2014, \aap, 567, A34

\bibitem[{{Wright} {et~al.}(2010){Wright}, {Eisenhardt}, {Mainzer}, {Ressler},
  {Cutri}, {Jarrett}, {Kirkpatrick}, {Padgett}, {McMillan}, {Skrutskie},
  {Stanford}, {Cohen}, {Walker}, {Mather}, {Leisawitz}, {Gautier}, {McLean},
  {Benford}, {Lonsdale}, {Blain}, {Mendez}, {Irace}, {Duval}, {Liu}, {Royer},
  {Heinrichsen}, {Howard}, {Shannon}, {Kendall}, {Walsh}, {Larsen}, {Cardon},
  {Schick}, {Schwalm}, {Abid}, {Fabinsky}, {Naes}, \&
  {Tsai}}]{2010AJ....140.1868W}
{Wright}, E.~L., {Eisenhardt}, P.~R.~M., {Mainzer}, A.~K., {et~al.} 2010, \aj,
  140, 1868

\bibitem[{{Wyatt}(2008)}]{2008ARA&A..46..339W}
{Wyatt}. 2008, ARA\&A, 46, 339

\end{thebibliography}



\renewcommand{\topfraction}{.5}
\onecolumn 

\begin{appendix}
\section{Multiple reductions of the \prim\ disk}
\label{sec:rdi}
\noindent
\begin{minipage}{\textwidth}
  \centering
  \includegraphics[trim={0 0 0 5cm},clip,width=16cm]{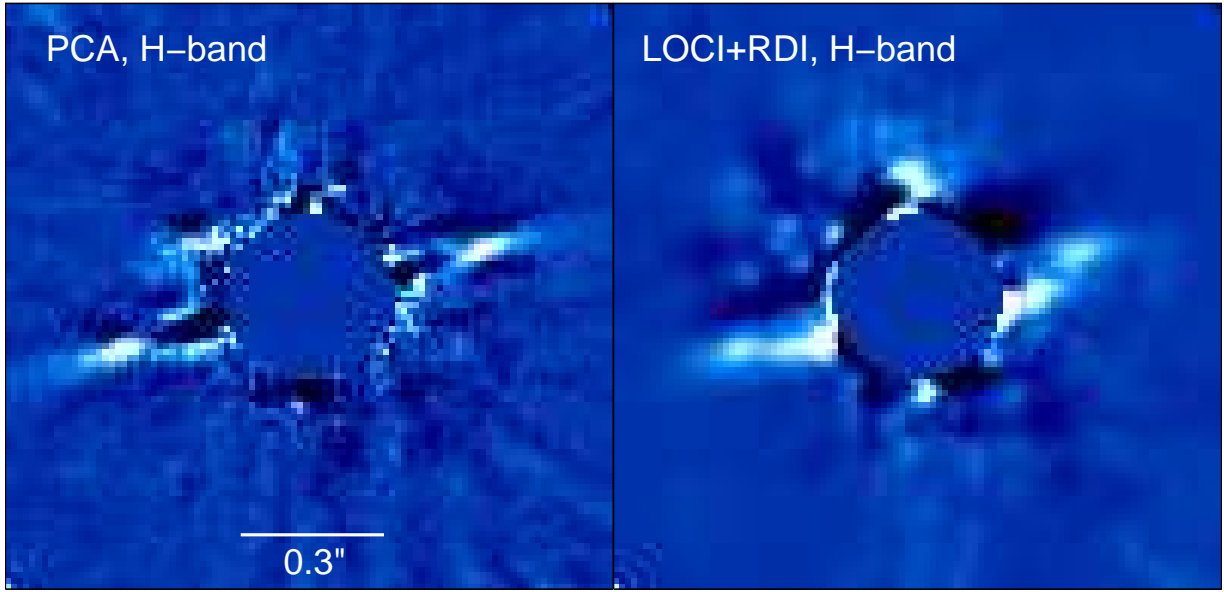}
  \captionof{figure}{Left: $H$-band PCA Reduced image of \prim . A highly inclined ring with large inner hole can be clearly
    seen. Right: RDI+LOCI reduction of same data set. The ring does not undergo self-subtraction here, and flux
    levels are more reliable.}
  \label{pca_rdiloci}
\end{minipage}

\vspace{10 mm}

\noindent
\begin{minipage}{\textwidth}
  \centering
  \includegraphics[trim={0 0 0 5cm},clip,width=16cm]{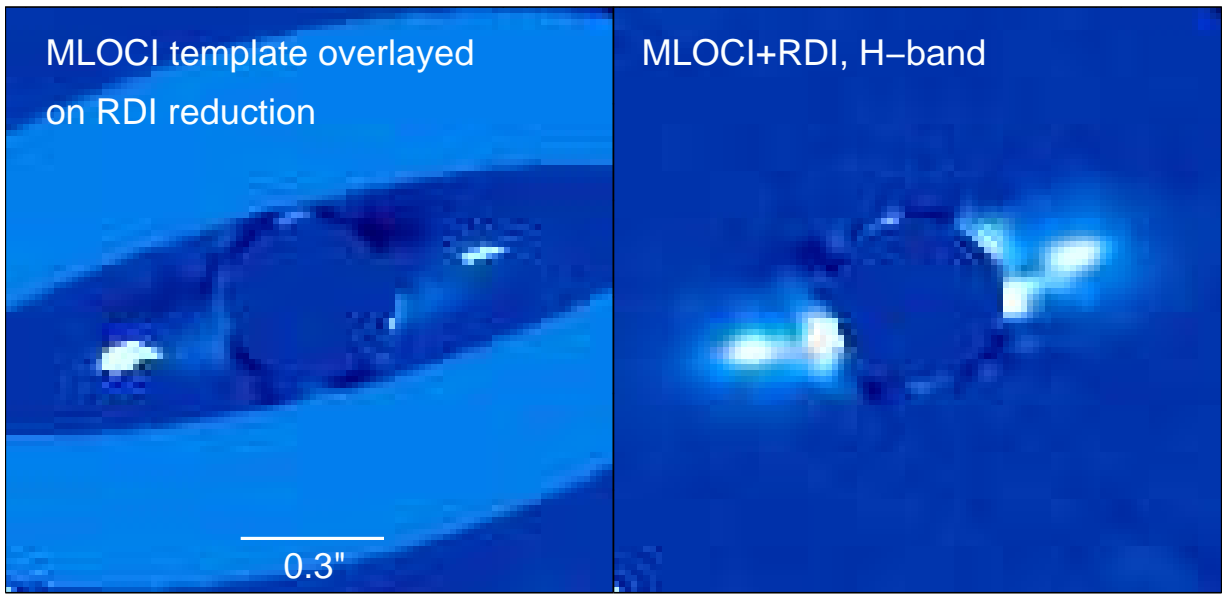}
  \captionof{figure}{Left: The MLOCI template. MLOCI works to preserve signal directly in the regions shown in white, while it tries to minimize noise in the solid blue elliptical region. 
    Right: MLOCI+RDI reduction using the template on the left. Same as in Figure~\ref{pca_rdiloci}, the ring does not
    undergo self-subtraction and the flux levels are more reliable.}
  \label{mloci_rdi_temp_red}
\end{minipage}

\vspace{10 mm}
Since the PCA reduction suffers from self-subtraction of the disk in
the star subtraction process \citep[see][]{2012A&A...545A.111M}, we need a higher fidelity image to
confirm the ring morphology.  The apparent sharpness of
the image is also due to this self-subtraction effect that enhances
the edges of the disk. We observed many stars without
any sign of a disk in the SHARDDS program, and so we can use these as a library
to construct reference images for star subtraction; this method is known as Reference
Difference Imaging (RDI).  The images from seven stars of similar brightness were
selected for the library; these are HD~10472, HD~105, HD~377, HD~25457, HD~38207, 
HD~206893, and HD~69830.  For each science image,
we chose the most similar images from this library, by comparing
the residual RMS in the annular region between 15 and 50 pixels from
the star, after they were scaled to minimize the residuals upon subtraction from the
science image. All pixels for which the disk emission is noticeable were
excluded from the match. 
Per science image, we selected a maximum of  60 library images,
with the condition that they reduced the RMS by  $>$30\%. At a minimum, we chose five of the best
matches. On average, the matches reduced the RMS by 38\% to 55\% (1\sig\ range). 
For star subtraction, we used the LOCI algorithm (Lafreniere et al.\ 2006) to subtract the best linear
combination of library (not science) images from each science image,  in annular rings of 20 pixels,
starting at 10 pixels separation from the star. 
The RDI reduced image, shown in Figure~\ref{pca_rdiloci}, thus has no
self-subtraction of the disk. We estimate the contrast of the disk in this image using unsaturated
images of the star, measuring 1.57$\times$10$^{-4}$ for the ring peak at the SE ansa.

We see  the ring with very similar extent and morphology as in the PCA reduction, albeit with a less
distinct inner hole. Although the disk is retrieved with higher fidelity,
here the removal of the star is not as effective.  Interestingly, we
identify a new feature of the disk: over the short axis of emission, one side of the disk is noticeably brighter. 
This could be due to either forward or backward scattering.

We also perform an MLOCI reduction using only the library images for
star subtraction (not images of the same star). We call this reduction MLOCI+RDI. 
The reduced image is shown in Figure~\ref{mloci_rdi_temp_red}. The star
subtraction is much more effective and a brightness asymmetry across the
short axis of the axis is seen again (see Section~\ref{sec:reduc}). 

\newpage

\section{Constraints on debris ring properties and \prim\ photometry}
\begin{table*}[!htp]
\small
\centering
\caption{Ring parameters estimates and 2$\sigma$ constraints from MCMC model comparisons. }
\begin{tabular}{llll}
\hline\hline 
Parameter                                               &       PCA Reduction                 & MLOCI+ADI Reduction         & MLOCI+RDI Reduction  \\\hline
X-offset (12.25 mas pixels)                   &      -1.1\ (0.0, -2.42)             &-1.2\  (-0.8, -1.7)                 &       -0.6\ (0.15, -1.3) \\                 
Y-offset (12.25 mas pixels)                   &      -1.3\ (-1.7, -0.6)              &-1.1 \ (-0.9, -1.16)               &       -0.2\ (-0.5, 0.42)     \\             
Inclination (\dg)                                     &      84.9\ (83.7,  86.9)             &83.3\ (82.9, 83.6)                 &          81.0\ (79.5, 81.9)   \\               
Position angle (\dg)                               &      -74\ (-73.2, -74.8)        &-74.3\ (-73.9, -74.6)           &        -73.6\ (-72.8, -74.3)  \\               
Inner radius ($r_{in}$, au)                        &      28.5\ (26.8, 30.4)              &27.6\ (26.9, 28.2)                  &           25.0\ (24.1, 25.8)  \\             
''Birth'' ring width ($\Delta r$, au)          &      5.0\ (1.8, 7.7)                    &1.9\ (1.04, 3.0)                       &               2.6\ (2.1, 3.12)   \\  
Outer density profile (exponent $\gamma$) & 4.8\ (3.4, 7.2)                    &3.9\ (3.6, 4.5)                         &          3.5\ (2.8, 3.9)   \\                          
Scattering anisotropy ($g$)                    &      0.07 (0.0, 0.18)                  &0.23\ (0.19, 0.27)                    &              0.23\  (0.15, 0.28)      \\    
Reduced $\chi^2$                                  &      1.8                                      &     1.5                                   &                2.7   \\
\hline
\label{tab:ringprams}
\end{tabular}
\end{table*}

\begin{table*}[!htp]
\small
\centering
\caption{Photometry and references.}
\begin{tabular}{lcccc}
\hline\hline 
Filter                                                       &      Central Wavelength                  & Flux (mJy)                          & Uncertainty (mJy)    & Reference  \\\hline
2MASS, Ks                                              &      2.159                                        & 909                                   &       28                    & \citet{2003tmc..book.....C}  \\                 
WISE, W2                                                &      4.6                                            & 239                                    &         8                     & \citet{2010AJ....140.1868W} \\ 
AKARI, MIR-S                                        &        9.0                                         & 104                                    &       10                    & \citet{2010AA...514A...1I}  \\             
MIPS24                                                   &      23.7                                      & 216.5                                 &         6                   & \citet{2014ApJS..211...25C} \\               
MIPS70                                                   &      71.4                                        & 350                                    &        36                    & \citet{2014ApJS..211...25C}  \\               
PACS100                                                 &      100                                           & 251                                    &      10                    &  This work \\             
PACS160                                                 &      160                                           & 111                                    &      30                    & This work  \\  
ALMA                            &         1240                                           & 0.43                                    &      0.05                    & \citet{2016ApJ...828...25L}\\  
\hline
\label{tab:phot}
\end{tabular}
\end{table*}

\section{Best SED model fits to \prim\ photometry}
\label{sec:sed}
\begin{figure}[ht]
  \centering
  \includegraphics[height=8cm]{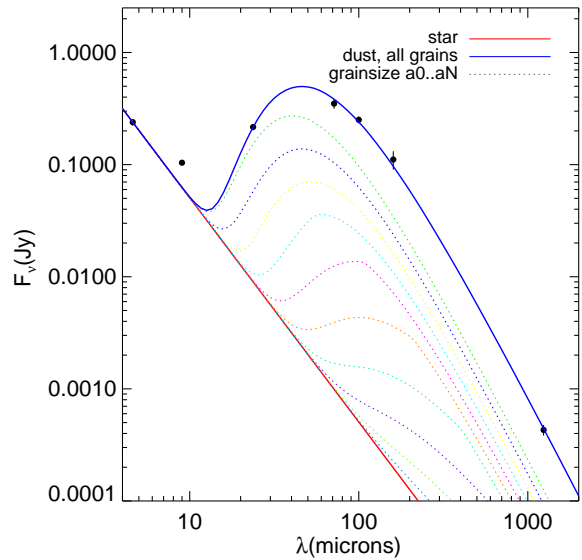}
  \caption{Photometry of \prim\ showing excess flux (black dots)
    over the photosphere (red line). The blue solid line is the best-fit model
    SED corresponding to the MLOCI+ADI model in Table~\ref{tab:ringprams} (disk inner radius of 27.6~au
    and $\gamma$=3.9; see text for details) with a  grain radius of 7.2~\mic\ and
    grain size power-law index, $q$=4.25. The dashed lines are the SED
    contributions of different grain diameters spaced evenly on a log scale (light green for 10~\mic , blue for 19~\mic , ... , purple for 1056~\mic).}
  \label{sed}
\end{figure} 

The photometry beyond 5~\mic\ where \prim\ has a detectible excess
flux over the photosphere is sparse. We
use the available flux measurements presented in Table~\ref{tab:phot} to constrain the dust properties of the system. 
The thermal emission at a particular wavelength from an annulus of infinitesimal width is modeled as 
$$f(r) = (1-\omega ) \Sigma(r) r^{-\gamma} \epsilon_{\lambda} B_{\nu}[T_p(a, r),\lambda](\frac{2\pi r dr}{D^2}) Jy.$$
See \citet{1992ApJ...385..670B} and \citet{2005ApJ...618..385W} for
details. Here, the new parameters are $\epsilon_{\lambda}$ ($=1.5a/\lambda$ when
$\lambda > 1.5a$, but 1 otherwise;\nocite{greenberg79}\ Greenberg 1979)  the radiative
efficiency,  $B_{\nu}$ the Planck function, $a$ the grain radius,
and $D$ the distance to the system in parsecs.  
The grain temperature for moderately absorbing grains 
like ''dirty ice'' (\eg,\ \nocite{greenberg79}Greenberg 1979) with radius $a$, is given by $T_p(r) = 432 (L_*/L_{\odot}/a)^{0.2} (r/au)^{-0.4}$.  
We assume that the grain size distribution is given by $n(a)~da \sim\
a^{-q}~da$ and the grain radius ranges from $a_0$ to
$a_1$(=2~mm). Here, $q=3.5$ would correspond to a steady-state
collisional cascade \citep{1969JGR....74.2531D}. When $a_1 \gg \lambda$,   the SED shape is
not very sensitive to $a_1$, since the number of large
grains fall so steeply both in number and temperature.
We compared our results to a quick run using the fitting tool SAnD \citep{2012A&A...541A.148E,2012A&A...537A.110L} and found consistent results from this independent tool and modeling approach.

\section{Detection limits on planetary companions}
\label{sec:conlim} 
There are many point sources ($>$ 20) detected in the IRDIS field of view
(11$''\times$11$''$), but given that this is a dense stellar field and that the candidate separations 
are large (\app 45~au), they are very likely background stars.
Follow-up observations will confirm or reject these as gravitationally bound
companions by testing for common proper motion. The 5$\sigma$
contrast from the PCA reduction was $\Delta H$= 9.0~mag at 0.25$''$ 
separation and 12.4~mag at 0.5$''$ separation. According to the
AMES-COND models \citep{2001ApJ...556..357A}, these detection limits
correspond to 38~\mjup\ and  16~\mjup\ at 0.25 and 0.5$''$ (or 23~au and
46~au), respectively.  These limits are much less stringent than the limits
we obtain from dynamical constraints considering the strength of
companion and disk interaction (see Section~\ref{sec:rdi}). 

\end{appendix}

\end{document}